\documentclass[preprint,amsmath,amssymb,
 aps,
 prl,floats]{revtex4-1}
\usepackage{graphicx}
\usepackage{dcolumn}
\usepackage{bm}
\begin{document}
\title{Atoms Talking to SQUIDs
\vspace{-6pt}}

\author{J. E. Hoffman$^1$, J. A. Grover$^1$, Z. Kim$^1$, A. K. Wood$^1$, J. R. Anderson$^1$, A. J. Dragt$^1$, M. Hafezi$^1$, C. J. Lobb$^1$,  L.~A.~Orozco$^1$, S. L. Rolston$^1$, J. M. Taylor$^2$, C. P. Vlahacos$^1$, F. C. Wellstood$^1$}
\address{$^1$Joint Quantum Institute, Department of Physics and National Institute of Standards and Technology, University of Maryland, 
College Park, MD 20742, United States.}
\address{$^2$Joint Quantum Institute, National Institute of Standards and Technology and Department of Physics, and University of Maryland, Gaithersburg, MD 20899, United States.}
\date{\today}

\begin{abstract}
We present a scheme  to couple trapped $^{87}$Rb atoms to a superconducting flux qubit through a magnetic dipole transition. We plan to trap atoms on the evanescent wave outside an ultrathin fiber to bring the atoms to less than 10 $\mu$m above the surface of the superconductor. This hybrid setup lends itself to probing sources of decoherence in superconducting qubits. Our current plan has the intermediate goal of coupling the atoms to a superconducting LC resonator.\\
\\
{\it{Hybrid quantum processor, neutral atoms, flux qubit }}


%
%
\end{abstract}

\maketitle 

\section{Introduction}

A hybrid quantum processor exploits the best aspects of its constituent qubits:  atomic systems exhibit excellent isolation from the environment, and
condensed matter systems 
benefit from technological developments in microfabrication, leading to a scalable qubit architecture\cite{Tian2004}. No known individual quantum system satisfies all the DiVincenzo criteria for a quantum computer \cite{DiVincenzo2000}; meeting these requirements is the driving force behind developing hybrid systems.

Here we propose a scheme to couple the ground state hyperfine transition of neutral $^{87}$Rb atoms to a superconducting (SC) flux qubit.  A flux qubit consists of three or more Josephson junctions in a micrometer-sized superconducting loop\cite{Friedman2000, Mooij1999, Robertson2006}. The supercurrent forms a superposition of clockwise and anticlockwise flow.  Typical designs involve one junction smaller than the others so that its Josephson inductance dominates the loop inductance.  We can use the microwave magnetic field from this SQUID loop to drive the atomic transition. The flux qubit possesses a long coherence time, on the order of 10~$\mu$s \cite{Nakamura2007} at the flux degeneracy point in a four-junction SQUID \cite{Yoshihara2006}, and a tunability  range of over a GHz\cite{Paauw2009}.

The atomic qubit consists of two magnetic sublevels on one of the hyperfine ground state manifolds of $^{87}$Rb, between the $| 5S_{1/2}; F=1  \rangle$ and $|5 S_{1/2};F=2 \rangle$ states, which are separated by about 6.83 GHz.

Similar proposals focus on creating a memory of ions \cite{Tian2004, Schuster2010a}, neutral atoms \cite{Sorensen2004, Verdu2009}, or molecules  \cite{Rabl2006, Andre2006, Schuster2010} to couple to a SC stripline resonator \cite{Imamoglu2009}.  Nitrogen-vacancy (NV) centers have taken a prominent role in these designs because of the ease with which these natural ion traps integrate with SC systems \cite{Marcos2010}.  Recent results demonstrate strong coupling of an ensemble of NV centers to a stripline resonator \cite{Kubo2010}.

\section{The Interaction and its Realization}
The flux qubit induces magnetic dipole transitions in the ground state hyperfine manifold of $^{87}$Rb.
The coupling occurs via the interaction Hamiltonian, $\sum_i -\boldsymbol{\mu}_i \cdot \boldsymbol B$, where $\boldsymbol{\mu}_i$ is the magnetic dipole moment of the $i$th individual atom, and $\boldsymbol{B}$ is the amplitude of the microwave magnetic field from the SC qubit at the same atomic frequency. To calculate the strength of the coupling we must first determine the strength of the magnetic field.

The average magnetic field associated with a single microwave photon is:
\begin{align}
B &= \sqrt{\frac{\mu_0 \hbar \omega}{2 V_{\mathrm{eff}}}}\, ,
\end{align}
where $\omega$ is the frequency of the photon, $\mu_0$ the permeability of free space, and $V_{\mathrm{eff}}$ the effective mode volume.  We take the effective mode volume for a square SQUID with 10~$\mu$m~$\times$~10~$\mu$m sides and width of 5~$\mu$m to be about $10\times 10^{-16}$~m$^3$, assuming the field is confined within 5 micrometers above and below the strip. For a photon at 6.8 GHz the magnetic field is about 1$\times 10^{-8}$ T, which for the typical moment of $1.4 \times 10^{10}$ Hz/T gives a coupling strength of roughly 100 Hz for one atom.

The field associated with a single quantum flux, $\Phi_0 = \frac{h}{2e}= 2*10^{-15} Tm^2$,
in a similar device corresponds to $2 *10^{-5} T$, a much higher value than the single photon.  The coupling of one magnetic flux to the atomic magnetic moment would be lowered by some geometric factor determined by the specific shape of the magnetic field obtaining a single flux coupling larger than 100 Hz. Further understanding of the field distribution and the inductance of the SQUID will help narrow the range of this number. 

We are working to first couple atoms to a high-Q lumped-element superconducting LC resonator, tuned to the $^{87}$Rb hyperfine splitting.
\begin{figure}[H]
\begin{center}
\includegraphics[width=0.9\linewidth]{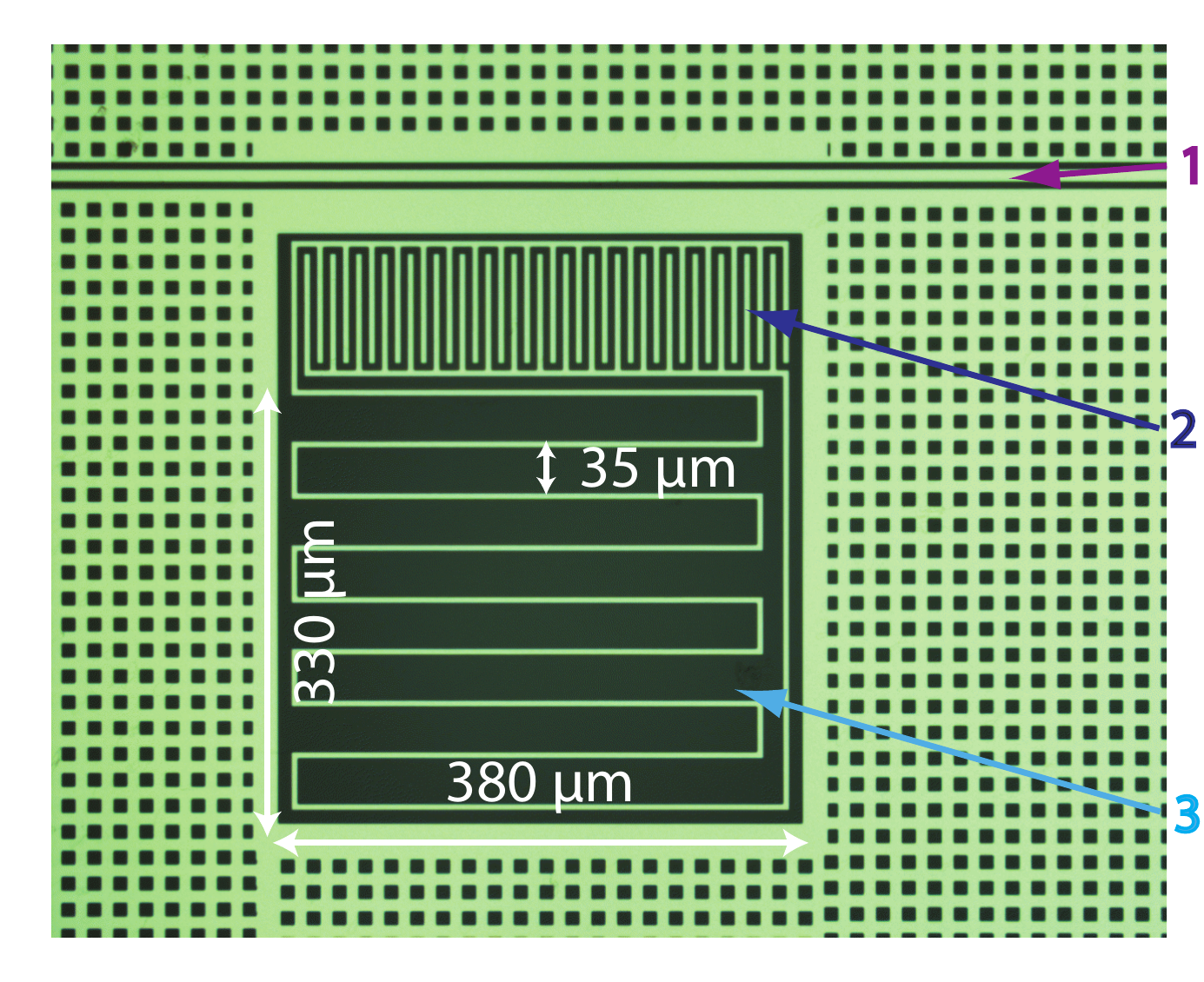} 
\caption{(Color online) Current lumped-element LC resonator.  1:  microwave transmission line. 2: interdigitated capacitor. 3: meandering inductor.}
\end{center}
\label{figresonator}
\end{figure}
Figure 1 presents our current lumped-element LC resonator. The niobium resonator consists of a meandering inductor and an interdigitated capacitor coupled to a transmission line. At a working temperature of 12 mK and on resonance at 6.863 GHz, the transmission through the microwave line decreases by 1.5 dB, and the loaded quality factor is 40,000. We simulate the electromagnetic fields of the SC resonator with the software package, High Frequency Structure Simulator (HFSS),
to determine the position of maximal magnetic field and uniformity.
\begin{figure}[H]
\begin{center}
\includegraphics[width= 0.9\linewidth]{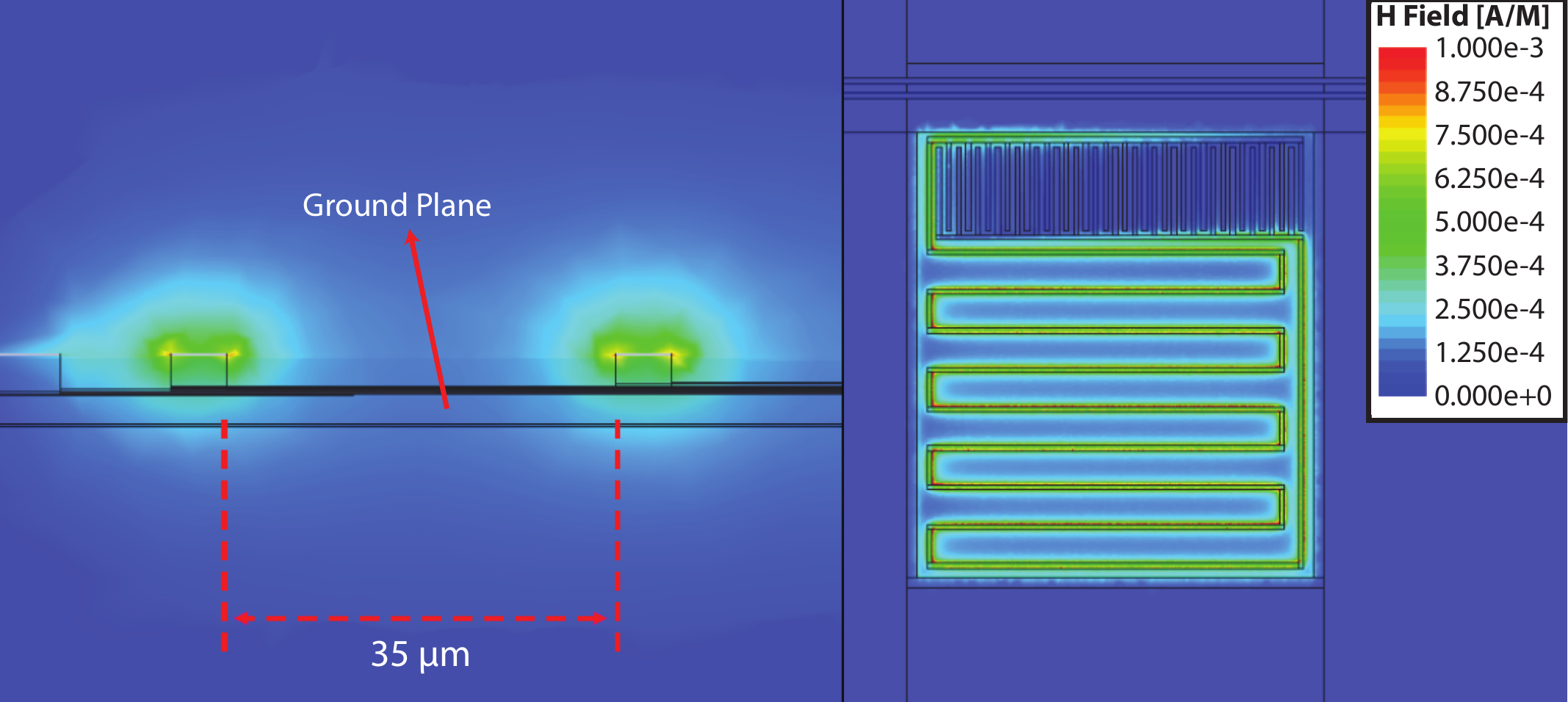}
\caption{(Color online) HFSS simulation of the H field produced by 0.016 photons in the lumped-element LC resonator.} 
\end{center}
\label{fig:HFSS}
\end{figure}
The coupling to the resonator depends on the specific geometry but will not reach the single photon value that we estimate for the SQUID. It will be lower by at least an order of magnitude. Figure 2 shows that the magnetic field outside of the SC as the field lines encircle the edges of the meandering inductor.  Using HFSS, we integrate the electromagnetic field over the entire simulation region to find the total energy with 
a total photon number of $n_{ph}$= 0.016 photons.  Then scaling the magnetic field located 5 $\mu$m above the inductor by 1/$\sqrt{n_{ph}}$ we obtain the magnetic field produced by one photon as $2.47 \cdot 10^{-9} T$.  We expect the coupling per atom to be 34.6 Hz.

\begin{figure}[H]
\begin{center}
\includegraphics[width=0.9\linewidth]{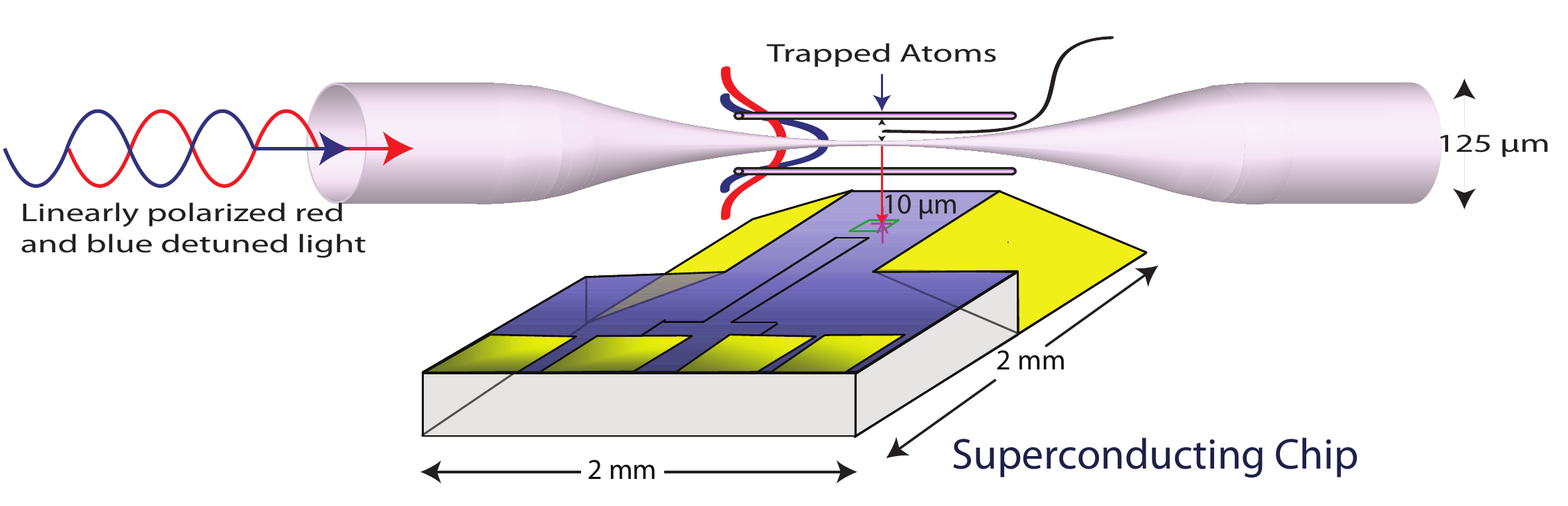} 
\caption{(Color online) A two color TOF trap hold atoms 10 $\mu$m above a flux qubit (small green square on the superconducting chip).}
\end{center}
\label{fig:FluxQbitTOF}
\end{figure}
The resonator operates inside a Triton 200 Cryofree Dilution Refrigerator from Oxford Instruments at 12 mK. The 200 $\mu$W cooling power of the final stage and degradation of the quality factor of SC resonators in a strong magnetic field present an unusual set of requirements for trapping atoms. These design constraints make the magneto-optical trap (MOT) and dipole trap difficult to implement \cite{Ovchinnikov1997, Metcalf1999}. 
We will realize an evanescent wave-based dipole trap outside an optical fiber whose diameter is smaller than the wavelength of input light. This trap is now called a tapered optical fiber (TOF) trap \cite{LeKien2004, Balykin2004, Vetsch2010} (see Fig. 3).

\section{The Superconducting Resonator}
\subsection{Tuning Scheme}
We have developed a frequency tuning system for a ``lumped-element'' thin-film superconducting Nb microwave
resonator on sapphire for coupling to cold $^{87}$Rb atoms.  The resonator must be tuned in order to match the resonance to the energy splitting of the ground state of $^{87}$Rb. We employ an Al pin as
a frequency tuner by placing it above the inductor and using a piezo-electrically driven stage to change the inductance of the resonator. 

\subsection{Tuning Performance}
Figure 4 displays the data for tuning the resonator. 
The present resonator has a large meandering inductor to ease alignment issues.

\begin{figure}[H]
\begin{center}
\includegraphics[width=0.9\linewidth]{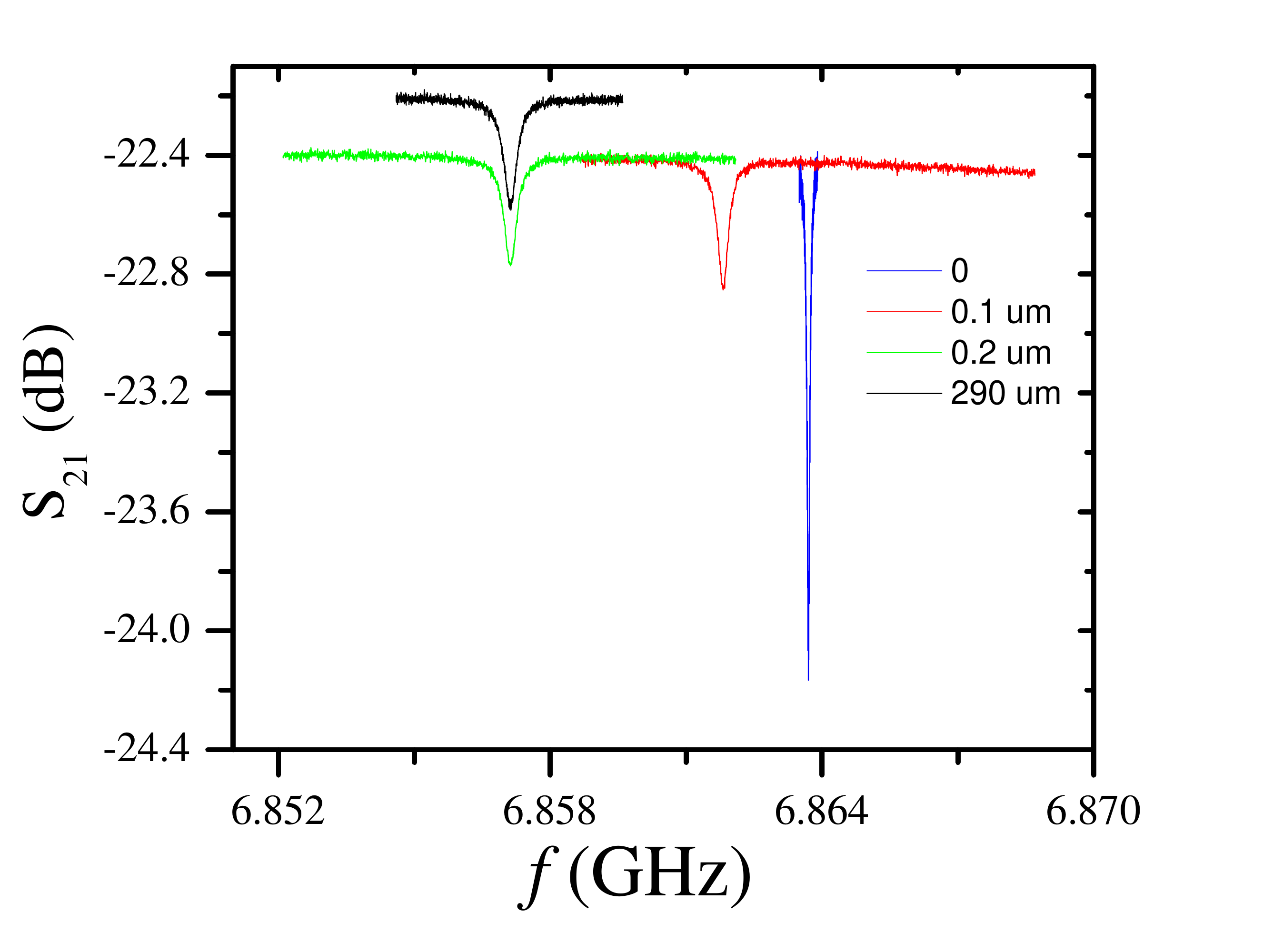} 
\caption{(Color online) The shift in the resonance frequency of the 
lumped-element resonator as a function of how far the aluminum pin moves. The blue, red, green, and black lines represent the aluminum tuning arm moving 0, 0.1, 0.2, and 290 $\mu$m, respectively.  These shifts yield resonances  of 6.8637, 6.8618, and 6.8571 GHz with loaded Q's of 87,000, 24,000, 19,600, and 18,700, respectively.}
\end{center}
\label{fig:tuning}
\end{figure}  

The data shows a decrease in the frequency as the tuning arm approaches the resonator (see Fig. 4). The first design allowed the pin to float as the piezo stage moves.  The ungrounded pin showed an unexpected shift of 66 MHz over a 290 $\mu$m range of motion, while lowering the loaded Q from 87,000 to 18,000.  
Connecting a copper foil from the pin to the sample box rectifies this grounding issue.

\section{Experimental Apparatus}
Figure 5 shows the current apparatus to test the LC resonator and effects of optical fibers on the resonator. 
\begin{figure}[H]
\begin{center}
\includegraphics[width=0.9\linewidth]{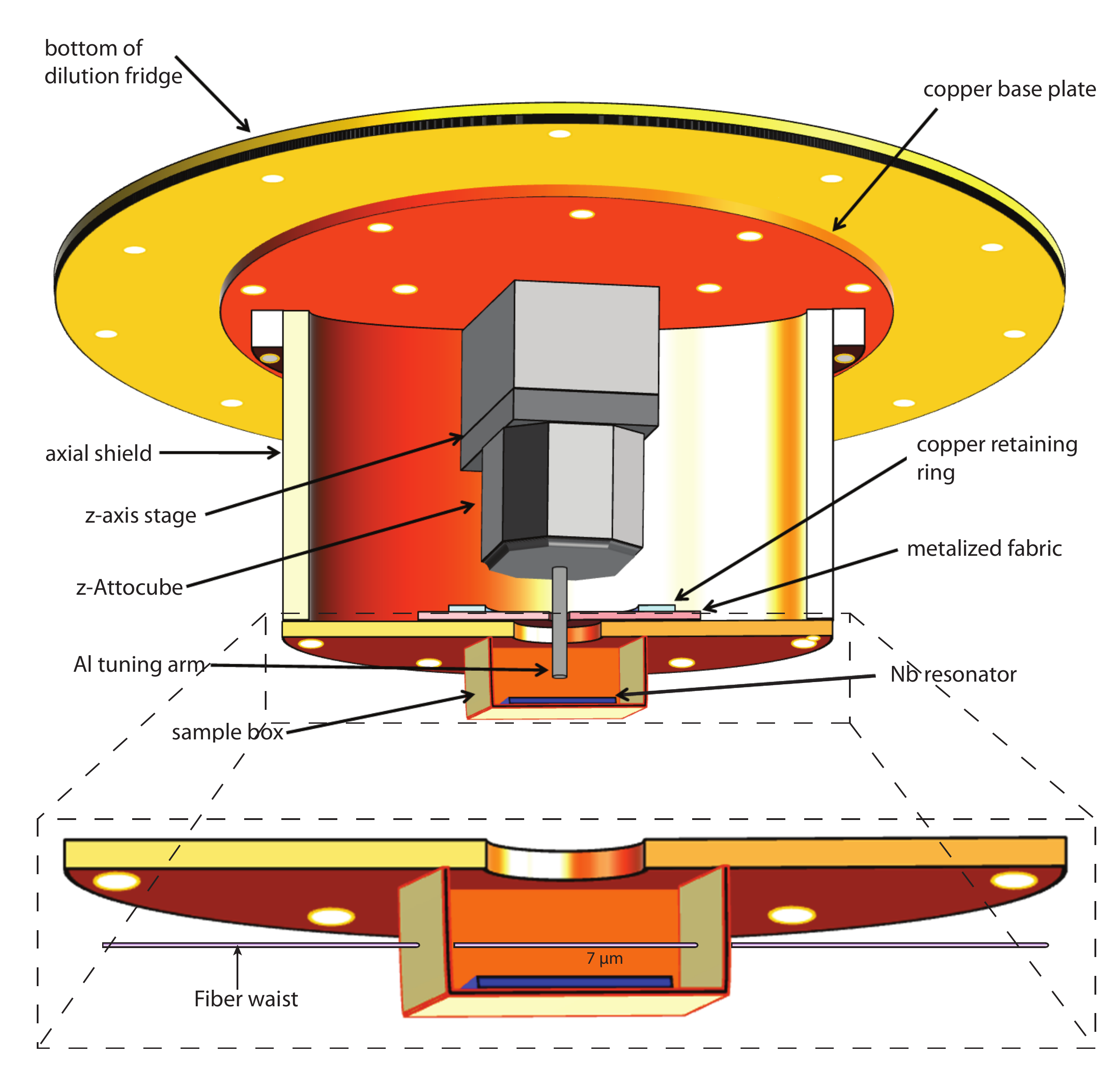} 
\caption{(Color online) The sample box anchored to the 12 mK stage of the dilution refrigerator.}
\end{center}
\label{fig:Fridge}
\end{figure}
It consists of a copper sample box thermally anchored to the bottom 12 mK stage of the refrigerator.

It contains a piezo-electric driven stage to adjust the position of a tuning wire above the LC resonator. We are currently designing the inital cooling scheme necessary to load atoms into the TOF trap. The ultrathin fiber would run through the cooled atoms allowing for their transfer to the TOF trap.  The TOF trap introduces the atoms into the sample box and is located 10 $\mu$m above the resonator providing for a coupling of about 10 Hz per atom.

\section{The Tapered Optical Fibers}
\subsection{Intensity Profile and Trapping Potential}
Figure 6 displays the intensity profile of the fundamental mode of a linearly y-polarized, 980 nm input laser about a 500 nm diameter fiber. Notice the discontinuity of the field when it leaks outside the glass fiber.
\begin{figure}[H]
\begin{center}
\includegraphics[width=0.9\linewidth]{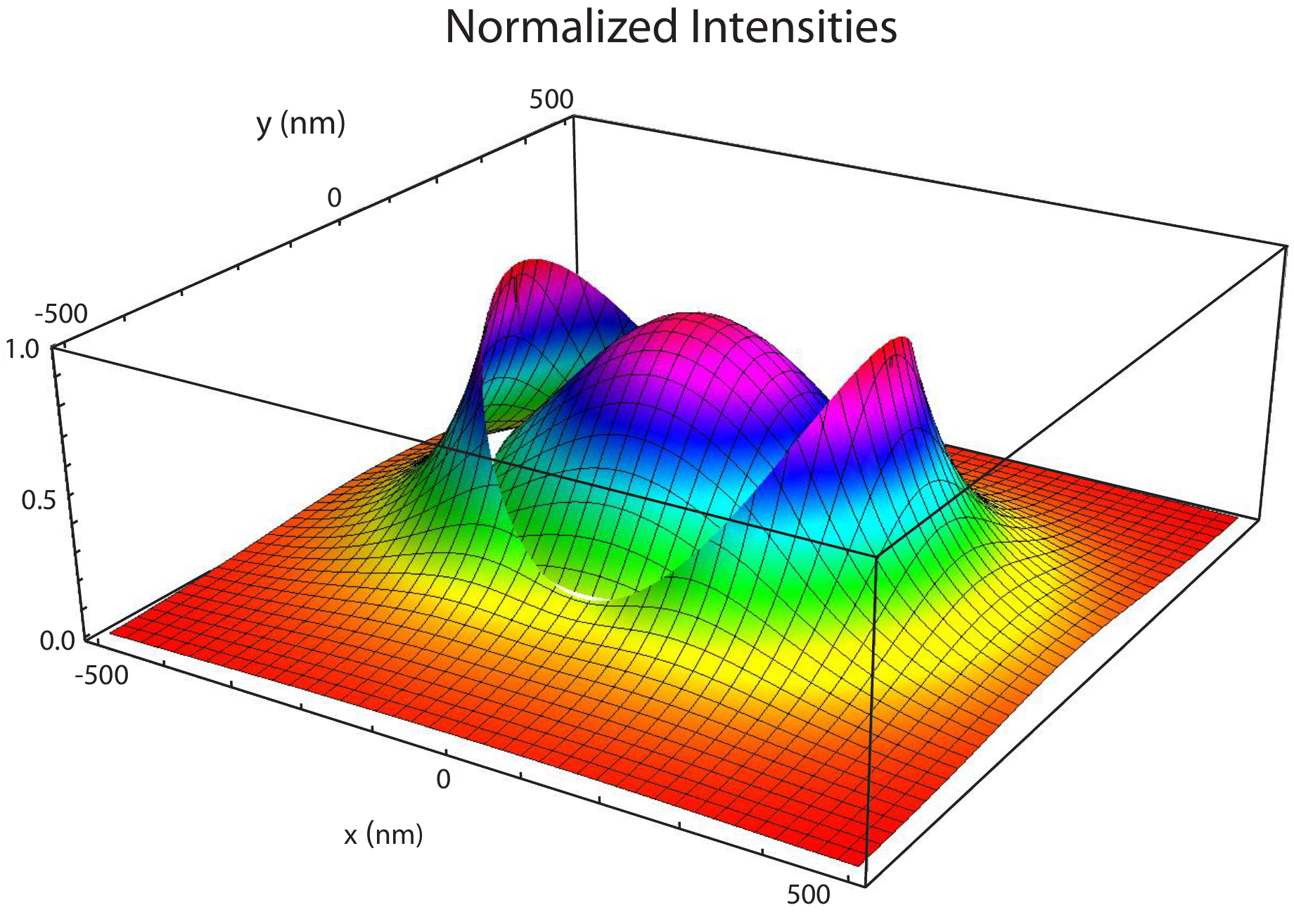} 
\caption{(Color online) Calculation of the intensity profile for a 500 nm fiber with an input of 980 nm linearly polarized light.}
\end{center}
\label{fig:EvWave}
\end{figure}
The intensity profiles inside and outside the fiber are \cite{Kien2004, Yariv1991}
\begin{align}
|E_{in}|^2 &= g_{in}\left[ J_0 ^2 ( h r) + u J_1 ^2 (h r) + f J_2 ^2 (h r)  \right. \nonumber \\ 
&\left.       + \left (u J_1 ^2 ( h r) - f_p J_0 ( h r) J_2 ( h r) \right)\cos\left[2\left( \phi - \phi_0\right)\right] \right] \\
|E_{out}|^2 &= g_{out}\left[ K_0 ^2 ( q r) + w K_1 ^2 (q r) + f K_2 ^2 (q r) \right. \nonumber \\
&\left.        - \left (w K_1 ^2 ( q r) - f_p K_0 ( q r) K_2 ( q r)\right) \cos\left[2\left( \phi - \phi_0\right)\right]  \right]  ,
\end{align}

where J$_n$ and K$_n$ are Bessel functions of the first and second kinds of order $n$ respectively, $r$ is the radius from the center of the fiber, and $g$ is a normalization constant.  The terms $h, q, f, f_p$ and $w$, given explicitly in \cite{Kien2004}, are functions dependent on the fiber radius and the propagation constant, $\beta$, of the input field. 

\begin{figure}[H]
\begin{center}
\includegraphics[width=0.9\linewidth]{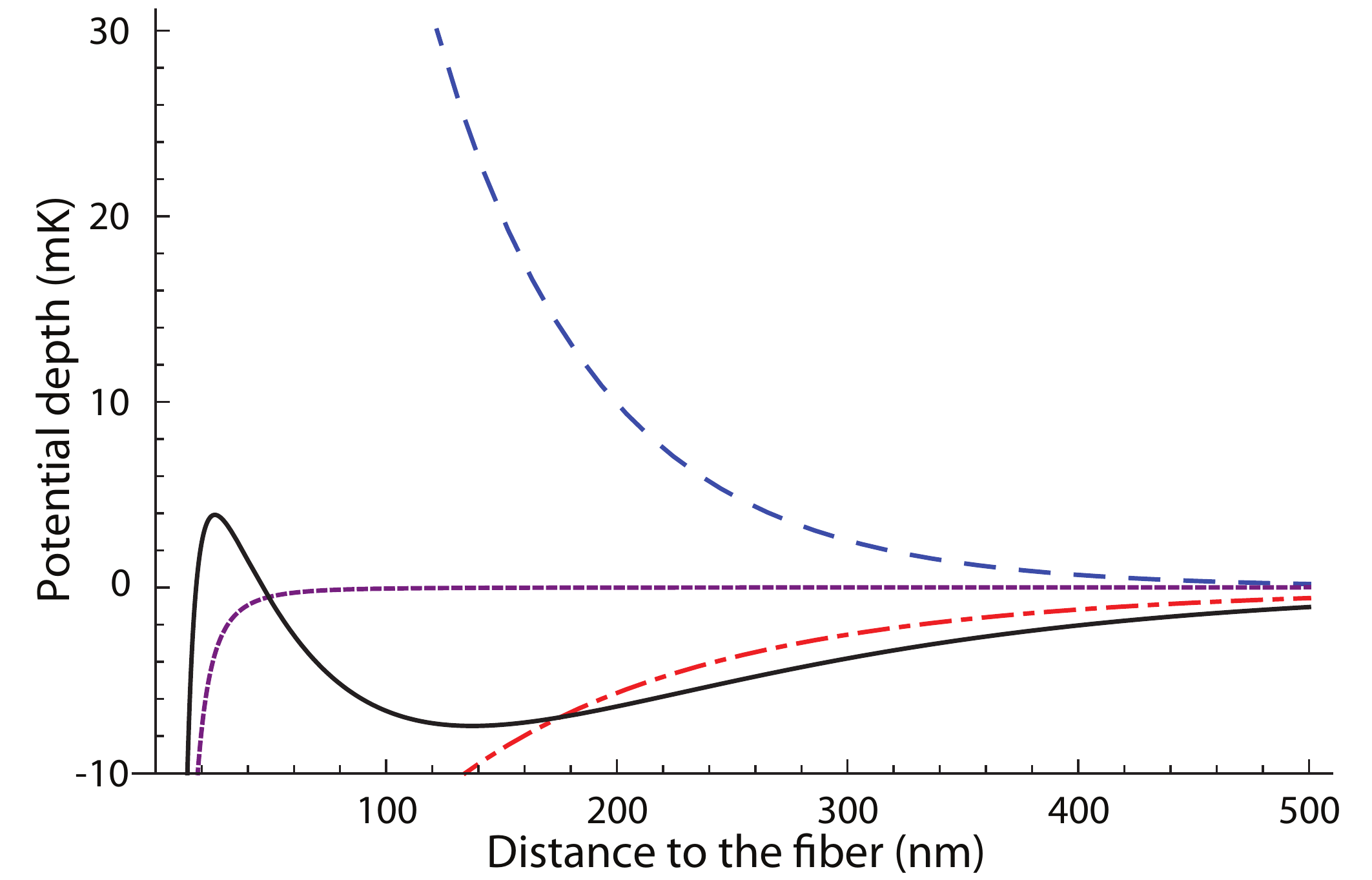} 
\caption{(Color online) The trapping potential for red- (long-short dashed line) and blue-detuned (dashed line) light of 980 and 730 nm with input powers of 30 mW and 13 mW, respectively.  The short dashed line represents the van der Waals interaction and the solid line is the total combined potential. These parameters yield a trap depth of 7.46 mK, 137 nm from the fiber surface.}
\end{center}
\label{fig:TrapPot}
\end{figure}

The intensity outside the fiber falls off exponentially with a characteristic decay length proportional to the wavelength of the laser light (see Fig. 6). We can realize a trapping potential by sending two color light, red- and blue-detuned from the D$_2$ transition of Rb, through the fiber \cite{LeKien2004}. The corresponding attractive and repulsive potentials combine to create a trap depth on the order of mK. 

Using linearly-polarized trapping light and coupling the red-detuned laser from both ends of the fiber, we will form 1D optical lattices along either side of the fiber waist. The lattice sites admit either zero or one atoms due to
collisional blockade effects, yielding $\sim$10,000 $^{87}$Rb atoms within a few hundred nm from the waist of a 500 nm TOF \cite{Schlosser2002}. Figure 7 shows a calculation of the trapping potential for a fiber with a 500 nm diameter. We apply lasers with wavelengths of 730 and 980 nm, and powers of 30 and 13 mW, respectively. These parameters produce a

trap depth of 7.46 mK with the atoms trapped 137 nm from the fiber surface. The ratio of red- to blue-detuned light controls the trap depth and position of the minimum. Increasing the red-detuned power, deepens the trap, and the atoms approach the fiber surface. 

The proximity of the atoms to the fiber surface means we must include 
the van der Waals interaction in the trapping potential. We do not, however, use the exact van der Waals interaction between an atom and a nanowire \cite{Boustimi2002} but instead treat the system as if the atoms were located next to an infinite dielectric, yielding a C$_3$ coefficient of 8.46$\times 10^{-49}$J$\cdot$ m$^3$ \cite{Courtois1996, Landragin1996}. Given the distances of atoms from the fiber surface, it appears that we may be in a cross-over region between the van der Waals and Casimir-Polder regimes.  To reconcile this issue we compare the asymptotic solution of the Casimir and Polder interaction \cite{Casimir1948,Zhou1995}, which accounts for retardation effects, to the dipole-dipole van der Waals term.

\begin{figure}[H]
\begin{center}
\includegraphics[width=0.9\linewidth]{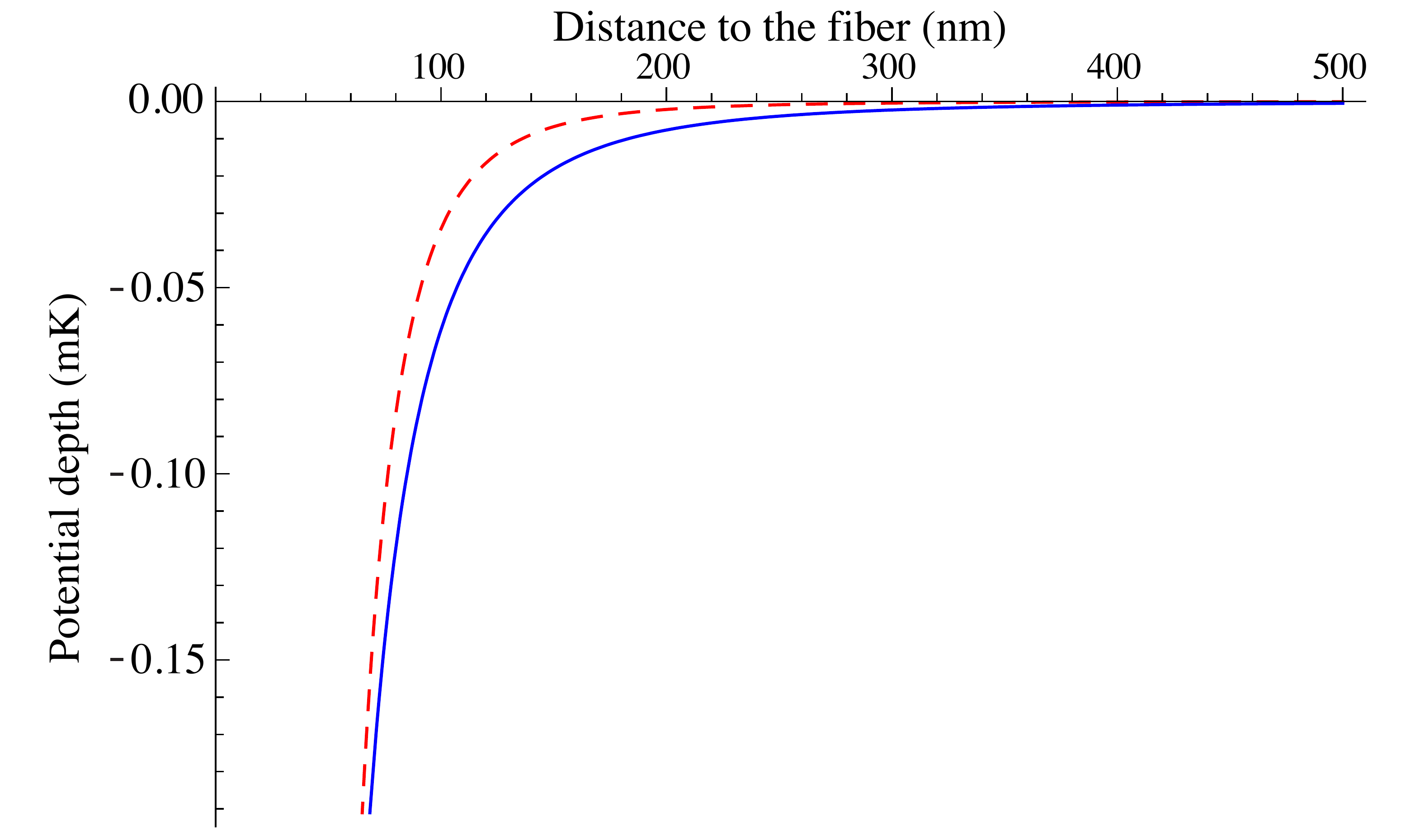}
\caption{The solid line represents the van der Waals potential while the dashed line denotes the Casimir-Polder.}
\end{center}
\label{fig:casimir}
\end{figure}
Figure 8 displays the van der Waals and Casimir-Polder potentials.  For this calculation we assumed a value of 0.0794$h$ $\,$Hz$\cdot$cm$^2$/V$^2$ for the ground state polarizability of Rb, and of 2.04 for the dielectric constant of silica, where $h$ is Plank's constant. Using Casimir-Polder, rather than van der Waals, we find a trap depth of 7.43 mK with intensity minima located at 138 nm.  This results in a shift in trap depth of 0.03 mK and atomic trapping sites of 1 nm, so we can neglect the Casimir-Polder interaction.
\subsection{Advances on the Fiber and its Realization}

The fiber pulling device necessary to construct a TOF requires a hydrogen-oxygen flame to create a local heat zone, while stepper motors pull the ends of a normal fiber until achieving the desired fiber waist radius. 
The TOF consists of three regions: the unmodified fiber, the taper or transition, and the fiber waist \cite{Sague2008}. How one modifies the length of the hot zone during the pulling process governs the shape of the TOF \cite{Birks1992}. The geometry of the transition region affects how light couples between the unmodified region and the fiber waist.    Poor coupling leads to scattered light, which heats the SC.  The design of the taper must satisfy an adiabatic criterion that best transfers light from the propagating mode in the unmodified region, LP$_{01}$, to the HE$_{11}$ mode in the fiber waist \cite{Love1986}.  Reference \cite{Love1991} gives the adiabatic condition as
\begin{align}
\Omega < \frac{\rho \left( \beta_1 - \beta_2 \right)}{2 \pi} ,
\label{eq:adiabatic}
\end{align}
where $\Omega$ is the angle of the taper, $\rho$ the radius of the fiber along the taper, and $\beta_{1,2}$  the propagation constants of the fundamental and first excited modes, respectively. When the light propagates in the unmodified region the core-cladding interface guides the mode, and the core contains a majority of the light.  As the fiber tapers, the core thins, and eventually the mode spreads into the cladding.  This radius change leads to a breaking of the single mode condition, and higher modes may be excited.  The radius of the transition region continues to decrease until the cladding-air interface guides the light.  When the cladding radius reaches a size that admits only a single mode, this acts as a mode filter, and all the higher modes scatter or reflect.  To avoid leaking light and heating the superconducting surfaces, the tapered region of the fiber must have a shape that satisfies the adiabatic condition of Eq. 4.
\section{Open Questions}
We are still investigating options to introduce the atoms to the TOF trap in the dilution refrigerator. Currently we plan to anchor a hollow copper rod to the 4K stage in the dilution refrigerator placed beside the sample box.  Inside the rod we will explore one of three options: a pyramid MOT \cite{Xu2008}, a diffraction grating-based pyramid MOT \cite{Vangeleyn2010}, or a trap relying on optical pumping \cite{Bouyer1994}.  The first two MOTs provide a compact design and require only a single laser.  The final option offers a trap without magnetic fields.  An electron beam driven atomic source, which requires only 10 mW of power \cite{Haslinger2010}, can introduce the atoms to the MOT.  The fiber waist runs through the center of the trap allowing for the transfer of atoms from the MOT to the TOF trap where an optical conveyor belt can move the atoms above the flux qubit or the LC resonator\cite{Kuhr2001}. 
\section{Conclusion}
We propose a scheme to couple $^{87}$Rb atoms, trapped along a TOF, to a flux qubit via the magnetic dipole moment. As an intermediate step the atoms will couple to a SC lumped element resonator. This project has applications in quantum information by working towards a hybrid quantum computer. Trapping atoms near a superconducting circuit also presents an opportunity to characterize sources of charge noise in the decoherence of SC qubits. Specifically, we can probe the source of charge noise by exploiting the atoms, or use Rydberg atoms, with their large dipole moments, as electric field sensors. 
This system can help to uncover the microscopic agents responsible for decoherence in SCs and use this to increase the coherence time. 
  
Work supported by National Science Foundation of the USA 
through the Physics Frontier Center at the Joint Quantum Institute.

\end{document}